# Statistical Mechanics Algorithm for Response to Targets (SMART)

## Lester Ingber

Ingber Physics Research and Physical Studies Institute
Drawer W, Solana Beach, CA 92075
and
Institute for Pure and Applied Physical Sciences
University of California at San Diego
La Jolla, CA 92093

## Abstract

It is proposed to apply modern methods of nonlinear nonequilibrium statistical mechanics to develop software algorithms that will optimally respond to targets within short response times with minimal computer resources. This Statistical Mechanics Algorithm for Response to Targets (SMART) can be developed with a view towards its future implementation into a hardwired Statistical Algorithm Multiprocessor (SAM) to enhance the efficiency and speed of response to targets (SMART_SAM).

## I. Identification of Sample Problems

*(A) Target Variables — Recognition*

Consider the grid in Fig. 1, defined within a given time epoch, where the grid is to be conceived as a generalized "radar" screen, representing data being accumulated by multiple sensors. Each cell has information pertaining to relocatable targets that may be moving between cells. This information is assumed to be further encoded into target variables, e.g., coordinate position, velocity, acceleration, numbers of targets within these categories, etc.

|  |  |  |
|---|---|---|
| × |  | × |
|  | × × |  |
| × × × | × | × × |

Figure 1. "Radar" Screen

The information collected within each time epoch serves to define changes in these variables between neighboring epochs, both within each cell and between neighboring cells.

Thus, large sets of problems are defined by requiring algorithms to recognize and parametrize changing patterns of these target variables.

*(B) Decision-Making Variables — Response*

It must also be assumed, if objective responses to targets are required, that decision-making variables be defined and functionally parametrized. These variables may include properties of actions to be taken, consistently scaled to match target variables.

Thus, larger sets of problems are defined by requiring algorithms to parameterize and to optimally allocate decision-making variables according to the perceived changing patterns of target variables defined in (A). It is also reasonable to expect that any algorithm for response, i.e., in contradistinction to mere recognition, somehow consistently fold in the parameters of both (A) and (B).

*(C) Response-Time and Computational Constraints*

These problems are further exasperated by the real nature of physical systems. Not much time may be available to optimally solve the problems defined in (A) and (B).

Thus, larger sets of problems are defined by requiring algorithms to respond to problems in (A) and (B), but so constrained that they may not be able to always predict the absolutely best response. It may be necessary to settle for a "good" response.

*(D) Fitting and Predicting Error, Noise and Risk*

Given the absence of perfect humans and of perfect machines, it is clear that any algorithm addressing the problems in (A), (B) and (C) require some degree of parametrization and modeling. There exist some errors in attempting to match any algorithm to a given genuine complex physical system. In order to minimize these errors to within required tolerances, these errors must be quantified.

By design of the targets or by design of the sensors, there also exists some degree of background noise tending to thwart a completely deterministic description of the target variables. This noise must be quantified, at least in order to assess a measure of credibility given to the identification of changing patterns of target variables.

The size and complexity of real physical systems, and the response-time and computational constraints described in (C), dictate that without always being able to make a best single decision, there exist elements of risk in any response algorithm. This risk must be quantified, at least in order to assess the chances to be taken by alternative responses. The "expected gain" of any response is the sum of products of each possible response multiplied by its associated risk, assuming independence among responses; otherwise, cross-correlations must be assessed and folded into this analysis.

Thus, larger sets of problems are defined by requiring algorithms to consistently include fits of variances (error, noise, risk) of all parameters in (A), (B) and (C). Only if variances are consistently fitted, can the mean values (signals), approximately corresponding to the otherwise deterministic parameters in the hypotheti-cal absence of these variances, be extracted. Only if past events include these "2nd moment" fits, i.e., only by fitting *bona fide* probability distributions, can the future be optimally predicted, albeit only with some (quantifiable) degree of statistical (un)certainty.

### (E) Introduction

Section II outlines the modern mathematical methods of nonlinear nonequilibrium statistical mechanics appropriate to the concerns of this section. Enough detail is given to establish the soundness of this approach, as well as to honestly present the algebraic complexity required.

Section III discusses applications of these mathematical methods to other specific physical systems. The intent is to demonstrate the broad applicability of this approach, as well as to offer some physical insights gained from related work that might be useful here.

Section IV suggests how this statistical mechanics approach might be applied in future projects, to develop computer software and hardware to solve problems of academic, industrial and governmental concerns.

## II. Method of Solution

### (A) One Variable, One Cell

For momentary simplicity, again consider the "radar" grid in Fig. 1, but now consider only one parameter, $M(t)$, in just one cell, representing just one of the variables discussed in Section (IA) or Section (IB). The problem of determining the change of $M$ within time $\Delta t$ is

$$M(t+\Delta t) - M(t) = \Delta t \, f[M(t)], \quad (1)$$

where $f[M]$ is some function to be fit, which describes how $M$ is changing. For small enough $\Delta t$, and assuming continuity of $M$, this is often written as

$$\dot{M} = \frac{dM}{dt} = f. \quad (2)$$

If background noise, $\eta$, is present, assumed to be Gaussian-Markovian ("white" noise), then this affects the description of changing $M$ by

$$\dot{M} = f + \hat{g}\eta, \quad (3)$$
$$<\eta(t)>_\eta = 0,$$
$$<\eta(t)\eta(t')>_\eta = \delta(t-t'),$$

where $\hat{g}^2$ is the (constant here) variance of the background noise. Here $\eta$ is assumed to have a zero mean. Eq. (3) is referred to as a Langevin rate-equation in the scientific literature.

Physicists and engineers, e.g., in fluid mechanics, recognize an equivalent "diffusion" equation to Eq. (3), defining a differential equation for the conditional probability distribution, $P[M(t+\Delta t)|M(t)]$, of finding $M$ at the time $t+\Delta t$, given its value at time $t$.

$$\frac{\partial P}{\partial t} = \frac{\partial(-fP)}{\partial M} + \frac{1}{2}\frac{\partial^2(\hat{g}^2 P)}{\partial M^2} \quad (4)$$

is known as a Fokker-Planck equation.

Some physicists, e.g., in elementary-particle physics, are familiar with yet another representation of Eq. (3) or (4). For small time epochs, the conditional probability $P$ is

$$P[M_{t+\Delta t}|M_t] = (2\pi\hat{g}^2\Delta t)^{-1/2}\exp(-\Delta t L), \quad (5)$$
$$L = (\dot{M}-f)^2/(2\hat{g}^2).$$

$L$ is defined to be the Lagrangian. This representation for $P$ permits a "global" path-integral description of the evolution of $P$ from time $t_0$ to a long time $t$, i.e., in contradistinction to the "local" differential Eq. (4). Labelling $u$ intermediate time epochs by $s$, i.e., $t_s = t_0 + s\Delta t$, in the limits $\lim_{u\to\infty}$ and $\lim_{\Delta t\to 0}$, and assuming $M_{t_0} = M(t_0)$ and $M_t = M(t \equiv t_{u+1})$ are fixed,

$$P[M_t|M_{t_0}] = \int \cdots \int dM_{t-\Delta t} dM_{t-2\Delta t} \cdots dM_{t_0+\Delta t} \quad (6)$$
$$\times P[M_t|M_{t-\Delta t}] P[M_{t-\Delta t}|M_{t-2\Delta t}]$$
$$\times \cdots P[M_{t_0+\Delta t}|M_{t_0}],$$

$$P[M_t|M_{t_0}] = \int \cdots \int \underline{D}M \exp(-\sum_{s=0}^{u}\Delta t L_s),$$

$$\underline{D}M = (2\pi\hat{g}_0^2\Delta t)^{-1/2}\prod_{s=1}^{u}(2\pi\hat{g}_s^2\Delta t)^{-1/2}dM_s,$$

$$\int dM_s \to \sum_{\alpha=1}^{N}\Delta M_{\alpha s}, \, M_0 = M_{t_0}, \, M_{u+1} = \dot{M}_t,$$

where $\alpha$ labels the range of N values of $M$. For notational simplicity, the indices $s$ and $\alpha$ often will be dropped in the following, but these time and range discretizations must of course be explicitly programmed in all actual numerical calculations.

There are at least three advantages to Eq. (6) over its equivalent representations, Eqs. (3) and (4): First, there is a variational principle defined by Eq. (6), wherein a set of Euler-Lagrange differential equations exist for $L$, directly yielding those values or trajectories of $M$ which give the largest contribution to the probability distribution $P$.

Second, because $P$ is a *bona fide* probability distribution, there exist Monte Carlo numerical algorithms for calculating Eq. (6), sampling the $M$-space without having to calculate all values of $M$ at all intermediate time epochs from $t_0$ to $t$ to find $P$. This numerical algorithm also has the nice feature of avoiding traps in local minima when there are deeper minima to be had, representing more probable states.

Third, the Lagrangian representation permits the use of importance-sampling techniques, efficiently giving more weight to more probable states.

### (B) Many Nonlinear Variables

It is possible to formulate Langevin equations generalized from Eq. (3),

$$\dot{M}^G = f^G + \hat{g}_i^G \eta^i, \quad (7)$$

$$i = 1, \cdots, \Xi,$$
$$G = 1, \cdots, \Theta,$$

where $G$ corresponds to any number of $\Theta$ variables, e.g., target and decision-making variables in (IA) and (IB), $f^G$ and $\hat{g}_i^G$ are arbitrarily nonlinear functions of any or all $M^G$, and of $t$, and the index $i$ corresponds to recognizing that there can be many different sources contributing to the variance of $M^G$. The Einstein summation convention is used for compactness, whereby any index appearing more than once among factors in any term is assumed to be summed over, unless otherwise indicated. The time of evaluation of $\hat{g}_{si}^G$ during $s$-epochs intermediate between $t_0$ and $t$, $\bar{t}_s$ between $t_s$ and $t_{s+1} = t_s + \Delta t$, must now be explicitly prescribed. Unless otherwise specified, a midpoint Stratonovich rule will be chosen here, using $M^G(\bar{t}_s) = \frac{1}{2}(M_{s+1}^G + M_s^G)$, $\dot{M}^G(\bar{t}_s) = (M_{s+1}^G - M_s^G)/\Delta t$, and $\bar{t}_s = t_s + \Delta t/2$. This choice is consistent with other physical systems, and allows the use of standard calculus.

The Fokker-Planck equation generalized from Eq. (4) is

$$\frac{\partial P}{\partial t} = VP + \frac{\partial(-g^G P)}{\partial M^G} + \frac{1}{2}\frac{\partial^2(g^{GG'}P)}{\partial M^G \partial M^{G'}}, \quad (8)$$

$$g^G = f^G + \frac{1}{2}\hat{g}_i^{G'}\frac{\partial \hat{g}_i^G}{\partial M^{G'}},$$

$$g^{GG'} = \hat{g}_i^G \hat{g}_i^{G'},$$

where a "potential" $V$ may be added, and is often useful to simulate boundary conditions. Note that the microscopic sources of the variances, indexed by $i$, are summed over and do not need to be fitted in this "mesoscopic" representation.

The path integral generalized from Eq. (6) is written as

$$P = \int \cdots \int \underline{D}M \exp(-\sum_{s=0}^{u}\Delta t L_s), \quad (9)$$

$$\underline{D}M = g_{0_+}^{1/2}(2\pi\Delta t)^{-1/2}\prod_{s=1}^{u}g_{s_+}^{1/2}\prod_{G=1}^{\Theta}(2\pi\Delta t)^{-1/2}dM_s^G,$$

$$\int dM_s^G \to \sum_{\alpha=1}^{N^G}\Delta M_{\alpha s}^G, M_0^G = M_{t_0}^G, M_{u+1}^G = M_t^G,$$

$$L = \frac{1}{2}(\dot{M}^G - h^G)g_{GG'}(\dot{M}^{G'} - h^{G'}) + \frac{1}{2}h^G_{;G} + R/6 - V,$$

$$[\cdots]_{,G} = \frac{\partial[\cdots]}{\partial M^G},$$

$$h^G = g^G - \frac{1}{2}g^{-1/2}(g^{1/2}g^{GG'})_{,G'},$$

$$g_{GG'} = (g^{GG'})^{-1},$$

$$g_s[M^G(\bar{t}_s),\bar{t}_s] = \det(g_{GG'})_s, g_{s_+} = g_s[M_{s+1}^G,\bar{t}_s],$$

$$h^G_{;G} = h^G_{,G} + \Gamma^F_{GF}h^G = g^{-1/2}(g^{1/2}h^G)_{,G},$$

$$\Gamma^F_{JK} \equiv g^{LF}[JK,L] = g^{LF}(g_{JL,K} + g_{KL,J} - g_{JK,L}),$$

$$R = g^{JL}R_{JL} = g^{JL}g^{JK}R_{FJKL},$$

$$R_{FJKL} = \frac{1}{2}(g_{FK,JL} - g_{JK,FL} - g_{FL,JK} + g_{JL,FK})$$
$$+ g_{MN}(\Gamma^M_{FK}\Gamma^N_{JL} - \Gamma^M_{FL}\Gamma^N_{JK}).$$

Note that the variance $g^{GG'}$ is the $GG'$-matrix inverse of the $G$-space metric $g_{GG'}$, $R$ is calculated to be the Riemannian curvature scalar, and $\Gamma^F_{JK}$ is the affine connection in this space.

*(C) Many Cells*

For many cells, i.e., $\Lambda$ cells indexed by $\nu$, the path integral in Eq. (9) is further generalized, essentially by expanding the parameter space from the set $\{G\}$ to the set $\{G, \nu\}$.

$$\tilde{P} = \int \cdots \int \underline{D}\tilde{M}\exp(-\sum_{s=0}^{u}\Delta t \tilde{L}_s), \quad (10)$$

$$\tilde{M} = \{M_s^{G\nu}|G=1,\cdots\Theta; \nu=1,\cdots,\Lambda; s=1,\cdots,u\},$$

$$\underline{D}\tilde{M} = \tilde{g}_{0_+}^{1/2}(2\pi\Delta t)^{-1/2}\prod_{s=1}^{u}\tilde{g}_{s_+}^{1/2}\prod_{G=1}^{\Theta}\prod_{\nu=1}^{\Lambda}(2\pi\Delta t)^{-1/2}dM_s^{G\nu},$$

$$\tilde{g}_s[M^{G\nu}(\bar{t}_s),\bar{t}_s] = \det(g_{GG'\nu\nu'})_s, \tilde{g}_{s_+} = \tilde{g}_s[M_{s+1}^{G\nu},\bar{t}_s],$$

$$\int dM_s^{G\nu} \to \sum_{\alpha=1}^{N^G}\Delta M_{\alpha s}^{G\nu}, M_0^{G\nu} = M_{t_0}^{G\nu}, M_{u+1}^{G\nu} = M_t^{G\nu},$$

$$\tilde{L} = \frac{1}{2}(\dot{M}^{G\nu} - h^{G\nu})g_{GG'\nu\nu'}(\dot{M}^{G'\nu'} - h^{G'\nu'})$$
$$+ \frac{1}{2}h^{G\nu}_{;G\nu} + \tilde{R}/6 - \tilde{V}.$$

Constraints may be placed on variables by adding them to the potential $\tilde{V}_s$, e.g., as $J_{sG\nu}M_s^{G\nu}$ with Lagrange multipliers $J_{sG\nu}$.

If a prepoint-discretization rule is adopted, transforming from the midpoint-discretized Feynman $\tilde{L}_s$ and $\tilde{g}_{s_+}$, to define $\dot{M}^{G\nu}(\bar{t}_s) = (M_{s+1}^{G\nu} - M_s^{G\nu})/\Delta t$, $M^{G\nu}(\bar{t}_s) = M_s^{G\nu}$, $\bar{t}_s = t_s$, and $\tilde{g}_{s_+} = \tilde{g}_s$, then a simpler expression is obtained for the Lagrangian, one in which the Riemannian terms are not explicitly present.

$$\tilde{L}' = \frac{1}{2}(\dot{M}^{G\nu} - g^{G\nu})g_{GG'\nu\nu'}(\dot{M}^{G'\nu'} - g^{G'\nu'}) - \tilde{V}. \quad (11)$$

However, although $\tilde{P}$ is invariant under this transformation, $\tilde{L}'$ does not possess the variational principle possessed by the Feynman Lagrangian $\tilde{L}$, so that if the prepoint-discretized $\tilde{L}'$ and $\tilde{g}_{s_+}$ are used to fit the data, then some tests must still be made to see how efficiently the path integral can be calculated using $\tilde{L}'$ instead of $\tilde{L}$ to globally scan the data.

*(D) Fitting the Information in the Lagrangian*

The quantity to be fit will be the information, i.e., the "entropy," $\tilde{I}$,

$$\tilde{I} = -\int d\tilde{M}_t \tilde{P}\ln(\tilde{P}/\underline{\tilde{P}}), \quad (12)$$

where $\underline{\tilde{P}}$ is a reference state. $\tilde{P}$ approximates the WKB most probable $\tilde{P}_{WKB}$, using $\tilde{L}_{WKB} = \tilde{L} - \tilde{R}/12$ with $\exp(-\sum_{s=0}^{u}\Delta t \tilde{R}/12)$ absorbed in $\underline{D}_{WKB}\tilde{M}$, but using $\tilde{L}$ has computational advantages, admitting the variational principle.

It is physically reasonable to assume, at least for a first-order fit to data, that a $\nu$-mesh can be chosen such that $f^{G\nu}$ and $\hat{g}_i^{G\nu}$, generalized from Eq. (7), depend functionally only only on nearest-neighbor (NN) interactions between cells, i.e., they are functions of $M^{G\nu}$ and $M^{G\nu_{NN}}$, where $\nu_{NN}$ represents NN cells to cell $\nu$. Also, it is reasonable to require that summing over $i$-sources of noise in $\hat{g}_i^{G\nu}\hat{g}_i^{G'\nu'}$ that contribute to $g^{GG'\nu\nu'}$ enforces NN correlations between $\nu$ and $\nu'$. (The deterministic aspect of longer ranged trajectories still can be modeled by the functional form given to $f^{G\nu}$, and both $f^{G\nu}$ and $\hat{g}_i^{G\nu}$ may depend explicitly on the time variable.) Then, in $\tilde{L}'$ in Eq. (11) all implied sums $\sum_\nu \sum_{\nu'}$ reduce to $\sum_\nu \sum_{\nu_{NN}}$, and $\tilde{L}'$ is limited to contributions from NN and next-nearest-neighbor (N$^2$N) interactions picked up from products of the means and variances. $\tilde{L}$ in Eq. (10) picks up higher order neighbors from the Riemannian terms, which must be kept in the absence of any other physical assumptions about the system. For some systems it may be reasonable to assume that only $\tilde{V}$ contains NN interactions in $\tilde{L}$ or $\tilde{L}'$, especially if the NN interactions are small enough to be expanded and separated out.

There is a price to pay for the conceptual simplicity achieved by Eq. (10), i.e., relatively complicated algebraic expressions, but this price is paid by coding the proper functional form of $\tilde{L}$. However, there is an important trade-off for the time of calculation. $\tilde{L}$ is a relatively large expression to be calculated, but the size of the calculation of $\tilde{P}$ goes up only as $\ln[\prod_{sG\nu} N^G]$, which is relatively very efficient for a large number of target and decision-making variables.

Eq. (10) [or first its equivalent prepoint discretization given in Eq. (11)] will be fit to the data by assuming functional forms for $\tilde{V}_s$, $g_s^{G\nu}$ and $g_s^{GG'\nu\nu'}$. The convergence of $\tilde{L}$ or $\tilde{L}'$ is expected to be quite good. I.e., even polynomial forms for $g_s^{G\nu}$ and $g_s^{GG'\nu\nu'}$, with coefficients to be fit, define a Padé approximate to $\tilde{L}$ usually giving better convergence than obtained for $g_s^{G\nu}$ or $g_s^{GG'\nu\nu'}$ separately. Also, note that $\tilde{L}_s$ is a single scalar function to be fit.

### III. Related Work

#### (A) Basic Mathematical Physics

Since the late 1970's, several mathematicians and physicists have developed previous techniques of functional integration of classical mechanical Fokker-Planck and quantum mechanical Schrödinger-type partial differential equations [1], to explore the statistical mechanics of quite general nonlinear nonequilibrium Gaussian Markovian systems—sometimes referred to as Langevin systems with "colored" or "multiplicative" Gaussian noise [2-16]. The development of a path-integral Lagrangian permits calculation of an explicit probability distribution describing the evolution/filtering of input patterns of information.

Furthermore, a true variational principle permits efficient examination of many properties of such systems, and the path integral directly presents itself for numerical calculations using importance-sampling Monte Carlo techniques [15, 17, 18]. Riemannian geometry [19] facilitates, but is not necessary, to derive these results.

From the Langevin equations, other models may be derived, such as the times-series model and the Kalman filter method of control theory. However, in the process of this transformation, the Markovian description typically is lost by projection onto a smaller state space [20, 21]. This work only considers multiplicative Gaussian noise. These methods are not conveniently used for other sources of noise, e.g., Poisson processes or Bernoulli processes. It remains to be seen if colored noise can emulate these processes in the empirical ranges of interest, in some reasonable limits [22]. For example, within limited ranges, log-normal distributions can approximate $1/f$ distributions, and Pareto-Lévy tails may be modeled as subordinated log-normal distributions with amplification mechanisms [23]. At this time, certainly the proper inclusion of colored noise, using parameters fit to data to model general sources of noise, is preferable to improper inclusion or exclusion of any noise.

It should be noted that while progress has been made in extending the application of the path-integral approach to abstract nonlinear nonequilibrium systems, there also has been much progress made in developing algorithms bringing other physical systems within range of the more conventional approaches of the equivalent Langevin equations, e.g., models of military conflict [24, 25], as well as of path-integrals directly, e.g., wave propagation [26-31] and optimization of wiring of computer chips [32]. However, I believe that the three systems outlined next—in (B), (C) and (D)—represent the first published works to require the application of the full nonlinear analysis.

#### (B) Neuroscience

I have detailed a theory of neocortical interactions which permits mesoscopic calculations of macroscopic activity based on microscopic interactions [33-40]. The rationale for this theory is at least partially justified by its results: a precise description of statistical influences on synaptic modification, including a model of long-term-memory and the dynamics underlying neuronal interactions, with detailed analytic and numerical support including Monte Carlo calculations [34, 35]; a derivation of the $7\pm 2$ rule of short-term-memory (STM) capacity yielding $7\pm 2$ statistically favorable information-states during observed time scales with properties reflecting the primacy versus recency effect [36, 37]; and calculations of macroscopic wave-like activity [38] consistent with other EEG studies and yielding propagation velocities consistent with movements of attention and hallucinations across the visual field.

This work may be considered a prototypical description of how an information-processing system may perform global scans of data seeking statistical matching of patterns, prior to or in parallel with more detailed local analyses.

### (C) Nuclear Physics

I also have applied modern methods of functional integration and statistical mechanics used to investigate neocortex, to quantum systems, discovering a new effect in velocity-dependent nuclear forces [41] contributing to the binding energy of nucleii [42-44].

This work describes how eigenfunction expansions [45] can be used to great advantage in matching patterns of highly nonlinear theoretical constructs with empirical data. This numerical problem is now relatively common in several disciplines, such as in nuclear physics [41, 46-49].

### (D) Economics

Additionally, I have applied these methods to the study of financial markets [50].

This work describes how to approach the "inverse" problem of constructing a predictive model from previous data. Here, the microscopic details at time scales required for action are insufficient to detail the functional form of the probability distribution at the coarser level required for analysis and decision-making.

In this context, I find it interesting that these mathematical methods appear to quite well model processing of patterned information gathered by a global attention similar to that required in describing neocortex, as discussed in (B) above, and as more commonly experienced in disciplines requiring the formulating of winning strategies within short time scales in situations of conflict and/or complexity [51-55].

### (E) Artificial Intelligence — Contrast and Compatibility

For the purposes of this study, it is appropriate to simplify the definition of Artificial Intelligence (AI), to include software and hardwiring algorithms designed to process information categorized by trees and loops of multiple indicators and patterns of these indicators. Sometimes these indicators are given graded weights to simulate uncertainty, without attempting to define *bona fide* probability distributions. Also, the enormous ranges of research and applications of AI methods, and the resulting software and hardwiring algorithms developed [56], may be inappropriate or too costly for many projects. E.g., much AI work pertaining to natural-knowledge interfacing and categorization of data bases is not relevant or useful for dynamic (changing in time) open (nonequilibrium) systems requiring short response-times and having some degree of unpredictability [57, 58].

However, SMART and AI methods may be complementary, useful for performing global and local searches, respectively, sequentially or simultaneously.

### (F) SMART Computer Hardwiring

As discussed in the next Section, SMART software using the theoretical and numerical developments described in the previous section may be implemented into SMART hardwiring, to design a Statistical Algorithm Multiprocessor (SAM). These SMART_SAM hardwiring requirements are somewhat different than those currently being implemented on parallel processors [59-62], and are modeled after my work in neuroscience discussed in (B) above.

## IV. Future Research and Development

### (A) SMART Software

An early part of sample data will be used to fit parameters to SMART, and then SMART will be tested, by comparing how well it does in predicting future alternative responses, relative to what actually transpires in the latter part of the data.

SMART can be integrated with some other AI-type algorithms, e.g., trees and loops of weighted indicators, to see if performance can be enhanced within required response times. For example, SMART can scan for multiple minima of $\bar{L}$, representing possible optimal patterns of response to targets, using relatively coarse meshes for time epochs and parameter increments during the fitting process. In comparison with results obtained by making this mesh finer, other AI-type search algorithms may do better within these restricted parameter ranges. Either SMART or the AI-type codes may call each other as subroutines to find optimal responses within required response-times.

### (B) SMART_SAM hardwiring

My work in neuroscience discussed above suggests an approach for implementing SMART into hardwiring, to design a Statistical Algorithm Multiprocessor (SAM).

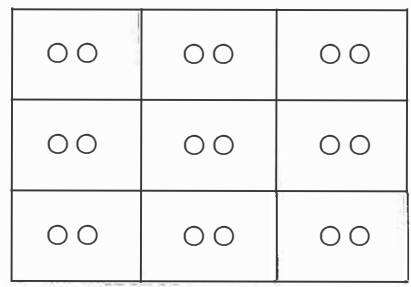

Figure 2. Statistical Algorithm Multiprocessor (SAM)

Consider Fig. 2 to have each cell of the "radar" screen in Fig. 1 now be represented as one $\nu$-cell of SMART_SAM at a given time labeled by $s$. Each circle consists of $\sim 10^2$ on-off bits, representing $N^G$ $\alpha$-states of one $G$-variable $M^{G\nu}_{\alpha s}$ in that $\nu$-cell at time $s$, which therefore represents a field rather than a simple binary node. Each circle statistically reacts to the other circles in that cell and in $\nu_{NN}$ cells at time $s-1$, according to an importance-sampling Monte Carlo algorithm encoded in each $\nu$-cell, e.g., as represented by Eq. (10) or (11). Long-ranged constraints might be added by superimposing (magnetic) fields, i.e., modeling the $J_{sG\nu}$ constraints described in Section (IIC) above.

*(C) Application to Other Systems*

It should be apparent that SMART and/or SAM can be used to solve pressing problems of industrial and governmental concern, ranging from adaptations for targeting and response systems, to developing sound probabilistic analyses aiding institutional investors in financial markets.